# Delegation Management Modeling in a Security Policy based Environment

Ryma Abassi

Higher School of Communication, SUP' COM,
University of Carthage
Tunis, Tunisia
ryma.abassi@supcom.rnu.tn

Sihem Guemara El Fatmi

Higher School of Communication, SUP' COM,
University of Carthage
Tunis, Tunisia
sihem.guemara@supcom.rnu.tn

**Abstract.** Security Policies (SP) constitute the core of communication networks protection infrastructures. It offers a set of rules allowing differentiating between legitimate actions and prohibited ones and consequently, associates each entity in the network with a set of permissions and privileges. Moreover, in today's technological society and to allow applications perpetuity, communication networks must support the collaboration between entities to face up any unavailability or flinching. This collaboration must be governed by security mechanisms according to the established permissions and privileges. Delegation is a common practice that is used to simplify the sharing of responsibilities and privileges. The delegation process in a SP environment can be implanted through the use of adequate formalisms and modeling. The main contribution of this paper is then, the proposition of a generic and formal modeling of delegation process. This modeling is based on three steps composing the delegation life cycle: negotiation used for delegation initiation, verification of the SP respect while delegating and revocation of an established delegation. Hence, we propose to deal with each step according to the main delegation characteristics and extend them by some new specificities.

## 1 Introduction

Nowadays, communication networks are more and more faced to the collaboration between entities. In order to avoid any unavailability or data loss, this collaboration must be secured and governed by involved entities permissions and privileges. In fact, any active entity, called subject, may need to share its permissions with another subject. This can be useful in case of an absence for example. Such kind of interaction is called delegation. Delegation is defined as the process whereby a user without any administrative prerogatives obtains the ability to grant some authorizations [1]. It can be considered as a potential approach in addressing the problem of providing dynamic access control decisions in activities with a high level of collaboration [2]. Hence, delegation is used in order to facilitate responsibilities interaction between subjects. The subject delegating a given permission is called grantor while the beneficiary is called, grantee. Moreover, a grantor can delegate some of his permissions to the grantee according to several characteristics including monotonicity (does the grantor preserve the delegated permission?), permanence (is the delegation permanent?), multiplicity (can several grantees benefit of a unique delegation?), level of delegation (can a grantee re-delegate an obtained permission?), lateral agreement (who is the delegation initiator?), etc.

In this paper, we propose a generic delegation model based on the most known delegation characteristics enumerated above. The proposed modeling is designed as an extension of a previously proposed framework using SP for network securing [7]. Consequently and for sake of compatibility, the present modeling is based on the same concepts such as subjects, channels, etc. and extends them by delegation's characteristics such as monotonicity, permanence, revocation, etc. A special attention is given to an important delegation aspect, revocation. Revocation is a significant aspect for delegation when permissions are required to get back. Since delegation can be made following several aspects, it seems obvious that its revocation also has to be made following several schemes.

The remaining part of this paper is organized as follows. In Section 2, we introduce our delegation modeling basing on several well established characteristics as well as new proposed ones. In Section 3, the whole delegation process modeling is depicted. Section 4 presents the revocation modeling and differentiates between several revocation schemas. Finally, Section 5 concludes this paper.

## 2   Related Work

During the last years, several works dealing with delegation have been proposed.

Barka and al. [4] proposed the first delegation model. Nevertheless, this model only deals with role level delegation, and does not consider sufficient constraints to manage the delegation policy. In [1], Ben Ghorbel-Talbi and al. describe a delegation approach for role-based access controls models. Their modeling is based on the AdOrBAC model and specifically on different administrative views. A view is an organizational concept used to structure the policy specification. Therefore, inserting an object in a view enable an authorized user to assign a user to a role and consequently to a permission. Conversely, deleting an object from a view, enable a user to perform a revocation. Recently, Crampton and al. [7], propose an orBAC model that focuses on the issue of roles and permissions transfer. In fact, this model defines specific components to manage each delegation level. For instance, predicates such as *grantP1* and *xferP1* are used to perform a permission delegation and a permission transfer, respectively. In [5], the authors focused on the delegation of obligations using the Alloy specification language.

Most of these works are based on Role-Based Access Control models and propose to extend it with some delegation aspects. Our proposition, in this paper, is a generic model that can be applied to several communication domains.

## 3   DELEGATION MODELING

In this section, we build our delegation modeling according to the delegation characteristics such as introduced in the literature.

### 3.1   Delegation basics

Let's recall that delegation is the process whereby a user without any administrative prerogatives obtains the ability to grant some authorizations such as introduced in [1]. The user initiating the delegation is called '*grantor*' while the delegation beneficiary is called '*grantee*'. In order to propose a complete framework dealing with delegation, we identified the main delegation characteristics [1, 4, 5, 7]. These latter comprise permanence, monotonicity, totality, levels of delegation, multiple delegation, lateral agreements and revocation.

*Monotonicity*: refers to the state of the authorization that the delegating member possesses after the delegation. A monotonic delegation means that upon delegation the delegating member maintains the delegated power. With a non-monotonic delegation, upon delegation the delegating member loses the delegated power for the duration of the delegation.

*Permanence*: refers to types of delegation in terms of their time duration.

*Levels of delegation*: defines whether or not each delegation can be further delegated and how many times. Single step delegation does not allow the delegation to be further delegated. Multi-step delegation allows the delegated member to further delegate his or her delegated permission to a third user, and so on.

*Lateral agreements*: refers to the delegation protocol between the delegator and the delegated members. It is of two types: bilateral agreement and unilateral agreement. A bilateral agreement is an agreement wherein delegation is accepted by both the delegating role member and the delegated member. A unilateral agreement, on the other hand, is a one-way decision.

*Totality*: in the context of role-based delegation models, totality refers to how completely the permissions assigned to that role are delegated.

*Multiple Delegations*: This type of delegation refers to the number of people to whom a delegating member can delegate at any given time.

*Revocation*: refers to the process by which a delegating user can take away the privileges that he or she delegated to another user who is a member of another role.

Based on the previously introduced characteristics, we depict in the following the proposed delegation modeling.

### 3.2   Monotonicity modeling

In this work, we distinguish between 'delegation' and 'delegation policy'. The first concept represents, as introduced above, the process allowing to a user to delegate some of his rights to another user. The delegation policy (DP) however, is a set of rules differentiating between legitimate and illegal delegations. Besides, we designate by "permission" any delegated right. A permission $p$ is structured similarly to a SP rule as follows:

$$p: \{type, modality, object, action, [constraint], [event]\}$$

Let's have a grantor $gr \in S$ (subjects set), a grantee $gt \in S$, $p$ a permission, some delegation constraints $dc \in C$ (constraint set), some delegation events $de \in E$ (event set) and $resp \in \{yes, no\}$, the response rule.

A grant request is formalized as follows:

$$grt\text{-}req\ (gr, gt, (p), dc) \rightarrow resp$$

A transfer request is formalized as follows:

$$tsf\text{-}req\ (gr, gt, (p), dc) \rightarrow resp$$

Unlike requests and according to the proof that we presented in [8], obligations cannot be granted but only transferred. Hence, we model a transfer obligation as:

$$tsf\text{-}ob\ (gr, gt, (p), dc, de)$$

Globally, a delegation rule can be represented by the following structure:

$$(t, m, gr, gt, p, dc, de)$$

where $t \in \{grt\text{-}req, tsf\text{-}req, tsf\text{-}ob\}$ is the rule type such as *grt-req* identifies a grant request; *tsf-req* identifies a transfer request and *tsf-ob* identifies a transfer obligation. $m$ is the rule modality, positive or negative. *gr* and *gt* represent respectively the grantor and the grantee. *p* is the delegated permission. *dc* are delegation constraints and *de* are delegation events.

### 3.3   Permanence and level of delegation modeling

Permanence is related to delegation constraints. Hence, we assume that a delegation is active while its constraints are valid which implies that constraints absence is synonym of a permanent delegation.

Furthermore, we propose to model three permanence constraint types: temporal constraints, spatial constraints and general constraints.

Temporal constraints '*t-c*' deal with the delegation duration. A temporal constraint can be defined through a given interval characterized by a begin date and an end date, or can occur before or after a given date. Furthermore, a temporal constraint can be a conjunction or a disjunction of other temporal constraints. For instance, the temporal constraint *BEFORE Wednesday* restricts the delegation use to Monday and Tuesday. This can be formalized as follows:

| | | |
|---|---|---|
| t-c | ::= | DURING interval | BEFORE date | AFTER date | ¬t-c | t-c ∧ t-c | t-c ∨ t-c |
| interval | ::= | '[' date '-' date ']' | literal |
| date | ::= | jj '/' mm '/' aa |
| jj | ::= | number |
| mm | ::= | number |
| aa | ::= | number |

The second constraint type that we consider deals with spatial constraints '*s-c*'. These latter restrict the application of the delegation process to a given site. So, we propose two constraint kinds related to the presence in a given site or the absence from it. This can be formally represented as follows:

| | | |
|---|---|---|
| s-c | ::= | IN location | ¬s-c | s-c ∧ s-c | s-c ∨ s-c |
| location | ::= | literal |

For instance, the spatial constraint *IN classroom* restricts the delegation use to the presence in the classroom.

General constraints '*g-c*' express varied constraints besides temporal and spatial ones related to the grantee such as his belonging, his old, etc. These constraints can be formally represented as follows.

| | | |
|---|---|---|
| g-c | ::= | HAS constraint | IS constraint | ¬g-c | g-c ∧ g-c | g-c ∨ g-c |
| constraint | ::= | literal |

Another essential characteristic for delegation is the delegation levels. It concerns the ability of the grantee to re-delegate a given delegation to another user. In this context, we propose another constraint type, the level constraints. Level constraints 'l-c' express a constraint authorizing delegated permission to be re-delegated. This means, that the grantee will be able to take advantage of the delegated permission or delegate it again. This is formally expressed as follows:

| | | |
|---|---|---|
| l-c | ::= | 'MULTI-LEVEL DELEGATION' |

### 3.4 Lateral Agreement Modeling

A delegation can be initiated spontaneously by the grantor or by a common agreement. In the first case, the grantor delegates his permission without referring to the grantee. Whereas, in the second case, the delegation can be initiated by the grantee who requests some permission from the grantor or by the grantor who delegates his permission after a common agreement with the grantee.

According to Fig. 1, three types of lateral agreement are modeled. In the first case, a spontaneous delegation is formulated. This can be modeled by the following elementary operations:
  (1) Grantor *gr* requests to delegate a permission to the grantee *gt* spontaneously and thus through a request submitted to the DP (*gr-write*)
  (2) The DP receives this requests (*DP-read*)

In the second case, the delegation is triggered by the grantee. This can be modeled as follows:
  (0-1') the grantee *gt* requests a given permission to the grantor *gr* (*dlg-claim*)
  (0-2') the grantor *gr* receives this request and accepts it (*dlg-approval*).
  (1) Grantor *gr* submits a request to the DP (*gr-write*)
  (2) The DP receives this requests (*DP-read*)

In the third case, delegation is triggered by the grantor. This can be modeled as follows:
  (0-1) The grantor gr request the grantee's approval for a delegation (*dlg-claim*)
  (0-2) The grantee gt receives this request and accepts it (*dlg-approval*)
  (1) The grantor *gr* submits a request to the DP (*gr-write*)
  (2) The DP receives this requests (*DP-read*)

Once the accord is achieved and the delegation process triggered, the DP can accept or deny this delegation. In the first case, the request will be handled according to its type i.e. transfer or grant where in the second case the request is simply dropped and a notification is sent back through steps (8) and (9) of Fig. 1.

### 3.5    Multiple Delegation Modeling

This characteristic refers to the number of grantees to whom a grantor can delegate the same permission at any given time. This number, let's say *Nd*, is fixed by the administrator. For each delegation request, the DP verifies that the number of delegations of the request's permission is less than *Nd*. This principle supposes the existence of a procedure counting the number of delegations each time a request, including the permission in question, is submitted by a grantor. We assume also, that the DP has a permission-Nd list from which it checks each arriving delegation request to decide whether it is bounded by a given number of delegation *Nd* or not.

Notice that when the request does not appear in this list i.e. Nd is not used, the number of grantees to whom a subject can delegate is not restricted.

## 4    Delegation process modeling

Based on the previously introduced concepts, our delegation modeling process can be depicted by Fig 1 as follows.
  (0-1)The grantor *gr* (respectively grantee *gt*) requests to delegate (respectively receive) a permission *p (dlg-claim)*.
  (0-2)The grantee *gt* (respectively grantor *gr*) accepts this delegation (*dlg-approval)*.
(1) The grantor *gr* puts into the channel the delegation request (*gr-write)*.
(2) The DP extracts this request (*DP-read*).
(3) The DP verifies the legitimacy of the request based on the existing SP (*dlg-legitimacy)*.
(4) If the request is accepted, a rule is added to the SP (*add-rule*).
  (4') In case of a transfer, the initial rule is simply modified (*modify-rule*)
(5) The DP then, notifies the grantee *gt* (*DP-write*)
(6) The grantee *gt* receives this notification (*gt-read)*.
(7) Nevertheless, if the request is denied, it is simply dropped into the channel *out (deny)*.
(8) A reject notification is then sent back to the grantor *gr (DP-write)*.

(9) The grantor gr receives the notification *(gr-read)*.

Let's note that the *dlg-claim* and *dlg-approval* steps concern the lateral agreement. Moreover, once accepted by the DP, a *grant* request implies the addition of a new rule in the SP. A *transfer* however, implies the modification of the rule handling the delegated permission (for the *grantor*) i.e. the *grantor* field is replaced by the *grantee* value.

Since, delegation updates an existing SP, a verification process must go with it. We propose two verification tasks: (1) delegation legitimacy verification when a delegation request is submitted and (2) SP consistency verification when the SP is updated.

Delegation legitimacy verification is performed by DP in order to check whether a given grantor has actually the right to delegate. Algorithm 1 depicts our proposition.

```
For each rule r ∈ SP \ (r.s==p.gr) do
    If (r∩ p <>∅) then
        If (r.modality == yes) then
            dlg-legitimacy
        else not(dlg-legitimacy)
    else not(dlg-legitimacy)
od
```

**Algorithm 1** Delegation request verification

Having a SP and a given permission *p* that a *grantor* wants to delegate to a *grantee*, this algorithm verifies the existence of a rule granting *p* for the considered *grantor*. Hence, we check only rules having a subject field equal to the *grantor*. For each of these rules, we compute its intersection with *p*. If the intersection is not empty then we look at the rule modality: a positive modality implies that the *grantor* has actually the delegated right where a negative modality implies that he hasn't this right. However, if the computed intersection is empty this mean that the *grantor* hasn't the permission *p* and consequently cannot perform the delegation.

The second verification task concerns SP consistency verification after an update due to delegation process. In fact, we have to verify whether the SP doesn't contain conflicts after the addition of the new rules. In a recent paper [9], we have proposed a framework detecting SP inconsistency. Hence, in the following we use the obtained results and interested reader may refer to the original paper for more details.

Having a SP (assumed consistent) and a given rule added by the delegation process *dr,* Algorithm 2 verifies whether there exist a rule *r* in the SP having a domain relation 'R' with the added rule *dr* and a contradictory modality. If such rule exists, then there is an inconsistency.

```
For each rule r ∈ SP \ (r.s == dr.s) do
        If (dom(r) R dom (dr)) and (r.modality != dr.modality)
        Then inconsistancy
        Else consistency
od
```

**Algorithm 2** SP consistency verification

Once a conflict is detected, it must be resolved. This resolution must take into account the fact that delegation is temporary process and that the initial SP must be preserved. Hence, we propose to re-solve conflict using priorities as well as partial order relation. Let's have P a finite set of priorities and let's assume that P is associated to a partial order relation < such that if $p_1$ and $p_2 \in$ P then $p_1 < p_2$ means that $p_2$ has a greater priority than $p_1$. We propose to associate to the rule added by the delegation process the greatest priority. In fact, delegation is performed in order to satisfy a given need in the SP environment consequently, a rule generated by delegation must have a greater priority than existing rules.

Furthermore, when the rule is revoked, then the partial order relation is deleted.

Let's consider the following example depicted two conflicting rules and where rule 1 was added by delegation:

Rule1: *(req, positive, assistant, present, course, during-professor-absence)*

Rule2: *(req, negative, assistant, present, course)*

These two rules cause a SP conflict since the SP will not be able to decide whether an assistant is authorized to present a course or not. Using our proposition, this is solved as follows. Two priorities are defined,

$P = \{p_1, p_2\}$ such that $p_1 < p_2$

Then the greater priority is associated to the rule added by delegation (Rule1).

Rule1: *(req, positive, assistant, present, course, during-professor-absence,* **$p_2$***)*

Rule2: *(req, negative, assistant, present, course,* **$p_1$***)*

Hence, Rule 2 will not be used by SP while delegation is not revoked.

## 5   REVOCATION MODELING

Delegation is needed, as we mentioned it, in order to facilitate user's interaction. This is the case, for example, when a user responsible of doing some tasks is not able to fulfill them. In such case, this user can delegate some of his permissions to another user. However, he must be able to recover them when initial constraints (of the delegation) are no more fulfilled. This is done by the revocation process. Revocation refers to the process by which a grantor can take away the permissions that he delegated to another user, the grantee.

In this section, we present our revocation framework dealing with its modeling and management.

### 5.1   Revocation dimensions

Following the classification defined by Hagstrom and al. [3], revocation can be categorized into three main different dimensions: *propagation*, *dominance* and *resilience*. However, while studying the main delegation features, we found that three other dimensions could be added: we propose *monotonicity*, *multiplicity* and *affiliation*.

**4.1.1. Propagation** distinguishes revocations according to space. Propagation is said *local* if the revocation affects only the direct grantees and *global* if its affects all grantees authorized by the direct ones [3]. A direct grantee is a grantee of the first level for the delegation.

In our modeling, this dimension is handled through two procedures *L-revoke* invoked when local propagation is used and *G-revoke* invoked when global propagation is used. Algorithm 3 depicts its principle: according to the request type, a new rule is removed (grant) or the grantee field is replaced by the grantor value (transfer).

---

**L-revoke** (p, gt, dp, sp)
  t := find (t,p,DP)
  r := find (p,sp)
  **if** (t == grt-req)
  **then** remove (r)      {grant}

```
      else   add-rule (r, gr) {transfer}
end
```

**Algorithm 3** Local Revocation Procedure

Algorithm 4 depicts the procedure *G-revoke* concerning the global propagation. For each grantee having received the permission from a direct grantee, we invoke the *L-revoke* procedure in order to remove (or modify) the concerned rule.

```
G-revoke (p, gt, dp, sp)
   L-revoke (p, gt, dp, sp)
   For all derived –dlg (gt') do
      L-revoke (p, gt', dp, sp)
   od
end
```

**Algorithm 4** Global Revocation Procedure

**4.1.2. Dominance** deals with conflicts arising when a subject losing permission in a revocation still has permissions from other grantors [3]. A revocation is said *strong* if the grantor initiating the revocation (revoker) dominates other grantors and revokes their delegated permissions too. It is said *weak*, however, when the revoker can only revoke permissions coming directly from him.

In our modeling, there is no restriction concerning the delegation of the same permission from several grantors. Hence, we adopt this dimension by defining a grantor hierarchy associated with a partial order relation 'dominates ⊑'. Formally, we consider H a finite set of hierarchies. We assume that H is associated with a partial order relation ⊑ such that if $h1 \sqsubseteq h2$ means that grantor $h1$ dominates $h2$ and consequently that the grantor associated with $h1$ dominates the grantor associated to $h2$.

**4.1.3. Resilience** distinguishes revocation via removal from revocation via a negative permission [3]. Therefore, a revocation is said to be *persistent* if a negative permission with a greater priority is given. The effect of this permission remains until it is revoked. The second case concerns the *non-persistent* revocation where the delegated permission is just removed.

In our modeling, we do not consider resilient revocation, since we considered that delegation concerns only positive permissions.

**4.1.4. Monotonicity** differentiates between the grant and the transfer delegation as introduced in section 4. Grant allows to a grantor to share some of his permissions while transfer allows to a grantor to hand over some of his permission to a grantee. Hence, we propose two revocation schemas, a monotonic revocation and a non-monotonic revocation, handling to the two types. A monotonic revocation (*delete*) concerns the grant revocation and induces the removal of the rule handling the concerned permission from the grantee(s). A non-monotonic revocation (*modify*) concerns the transfer revocation and induces simply the modification of the rule (handling the concerned permission for the grantee) by replacing the subject field by the grantor value.

**4.1.5. Multiplicity** deals with multiple revocations. It is said *multiple* if the delegation affects all the grantees of a given delegation and *single* if the revocation concerns only one grantee. We have to note, however, that the loss of the permission by the grantor implies an automatic revocation of all his grantees.

**4.1.6. Affiliation** deals with revocation triggers. In fact, we found that a revocation can be triggered according to three cases: (1) a constraint violation e.g. an expired date (2) a grantor request or (3) the loss by the grantor of the permission. If the grantor loses a permission he delegated, it is obvious that the delegation must be revoked.

These dimensions are combined to provide sixteen revocation schemes. However, in the following sub-sections, we describe only four schemes due to the paper length limits.

### 5.2   Revocation categorization

We have to note, that only the grantor request affiliation dimension can be combined with other dimensions. In fact, the two other (the constraint violation and the grantor permission loss) trigger an automatic revocation. In the first case, only the grantee (s) related to the constraint is (are) affected while in the second case, all direct grantees are affected (and consequently their direct grantees, etc.). Table 1 depicts the remaining schemes where each one has a unique description with respect to the four dimensions.

**4.2.1. Weak local single delete.** In this schema, there is no dominance and no propagation, only one grantee is affected and the rule issued from the delegation is deleted.

Let's consider the situation where an assistant professor 'A' belongs to several university departments. When 'A' no longer belongs to a given department, professor 'P' revokes A's authorizations, but he may still have access to the same objects as a member of other departments. Other assistants are not affected except that their direct grantor may no longer be 'A'. For 'P' to weakly locally single revoke a permission given to 'A', the operation consists in removing one permission and making sure that all the permissions 'A' has granted are still valid. If necessary, 'P' must assume the grantor role for the permissions that 'A' granted.

**4.2.2. Weak local plural delete.** This schema differs from the single variant in the plurality aspect, i.e. the professor 'P' attempt to revoke a permission granted to a given assistant 'A' will remove also in a ripple effect the other grantees beneficiary of the same permission from the same 'P'.

**4.2.3. Weak global single delete.** This schema is useful if professor 'P' loses trust in assistant 'A' but still trusts others grantees having received the same permission at the same time. Also, since 'P' no longer trusts 'A' in term, he no longer trusts any subject trusted by 'A'.

**4.2.4. Strong local plural delete.** In this schema, the revocation affects all direct grantees whatever their grantor.

## 6   Conclusion

The importance of security in communication networks is no longer in question. Hence, any efficient security solution must be adapted to the specific needs of the network that it governs. One of these needs concerns users' interaction where a given user may share its permissions or even transfer them. This is what we call delegation.

In this work, we proposed a delegation modeling designed to complete a previously proposed SP modeling and to interact with him. Hence, we separate two delegation types: the grant and the transfer and for each type, we depict a modeling dealing with the initial agreement process, the permanence, the delegation levels, multiple delegations and the revocation. Three agreements processes were proposed based on the initiator of the delegation. We also formalized three request rules: grant request, transfer request and revocation request. Moreover, a special focus was given to revocation management and thus through five dimensions: two preexist i.e. propagation and dominance and three newly proposed i.e. monotonicity, affiliation and plurality. These dimensions were combined to provide six-teen revocation schemes. Finally, we proposed a formal syntax formalizing our modeling. This syntax will be the first step towards implementation.

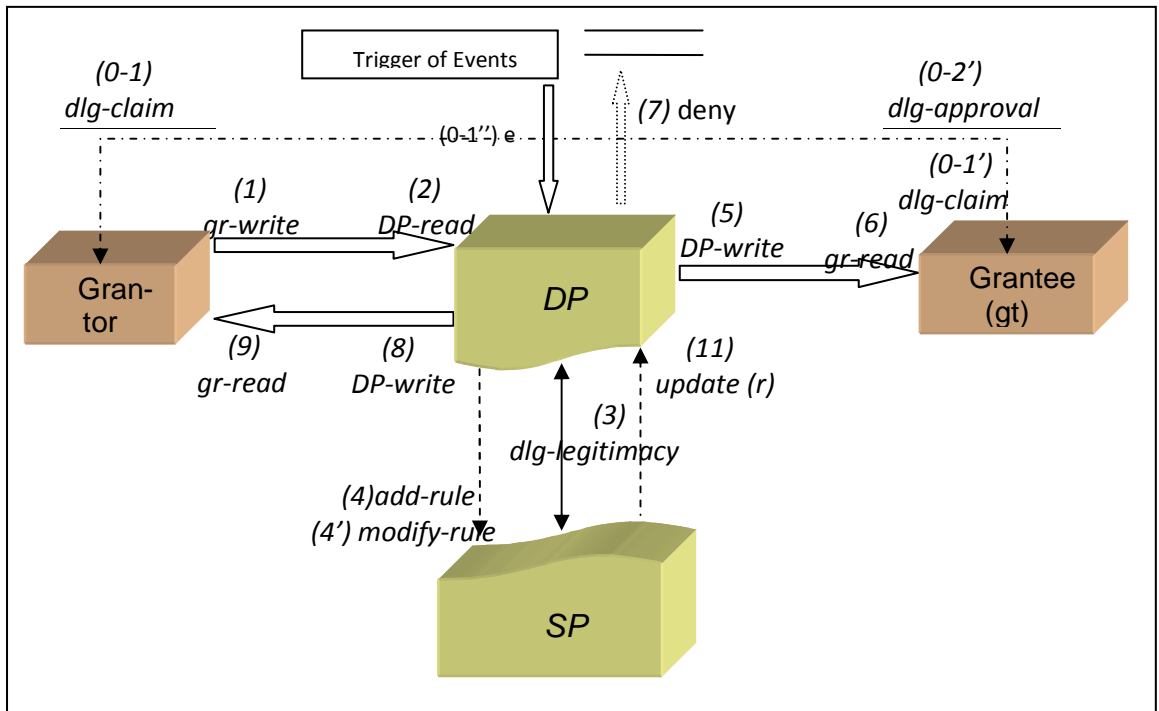

**Fig1**. The Generic Delegation Model

| Affiliation | Monotonicity | Dominance | Propagation | Plurality | |
|---|---|---|---|---|---|
| Grantor request | No | No | No | No | weak local single modify |
| | No | No | No | Yes | weak local plural modify |
| | No | No | Yes | No | weak global single modify |
| | No | No | Yes | Yes | weak global plural modify |
| | No | Yes | No | No | strong local single modify |
| | No | Yes | No | Yes | strong local plural modify |
| | No | Yes | Yes | No | strong global single modify |
| | No | Yes | Yes | Yes | strong global plural modify |
| | Yes | No | No | No | weak local single delete |
| | Yes | No | No | Yes | weak local plural delete |
| | Yes | No | Yes | No | weak global single delete |
| | Yes | No | Yes | Yes | weak global plural delete |
| | Yes | Yes | No | No | strong local single delete |

**Table 1.** Categorization of revocation schemes